\documentclass[mathleft,fleqn,%
]{an}
\usepackage{graphicx}

\begin{document}

\Pagespan{1}{}
\Yearpublication{2014}%
\Yearsubmission{2014}%
\Month{0}%
\Volume{999}%
\Issue{0}%
\DOI{asna.201400000}%

\renewcommand{\bf}{}

\title{Radial migration in numerical simulations of Milky Way-sized galaxies}

\author{R.\,J.\,J. Grand\inst{1,2}\fnmsep\thanks{Corresponding author:
        {robert.grand@h-its.org}}
\and  D. Kawata\inst{3} 
}
\titlerunning{Galactic spiral arms and radial migration}
\authorrunning{R.\,J.\,J. Grand \& D. Kawata}
\institute{
Heidelberger Institut f¡¯ur Theoretische Studien, Schloss-Wolfsbrunnenweg 35, 69118 Heidelberg, Germany
\and 
Zentrum f¡¯ur Astronomie der Universit¡¯at Heidelberg, Astronomisches Recheninstitut, M¡¯onchhofstr. 12-14, 69120 Heidelberg, Germany
\and 
Mullard Space Science Laboratory, University College London, Holmbury St. Mary, Dorking, Surrey, RH5 6NT, United Kingdom}

\received{XXXX}
\accepted{XXXX}
\publonline{XXXX}


\abstract{We show that in $N$-body simulations of isolated spiral discs, spiral arms appear to transient, recurring features that co-rotate with the stellar disc stars at all radii. As a consequence, stars around the spiral arm continually feel a tangential force from the spiral and gain/lose angular momentum at all radii where spiral structure exists, without gaining significant amounts of random energy. We demonstrate that the ubiquitous radial migration in these simulations can be seen as outward (inward) systematic streaming motions along the trailing (leading) side of the spiral arms. We characterise these spiral induced peculiar motions and compare with those of the Milky Way obtained from APOGEE red clump data. We find that transient, co-rotating spiral arms are consistent with the data, in contrast with density wave-like spirals which are qualitatively inconsistent. In addition, we show that, in our simulations, radial migration does not change the radial metallicity gradient significantly, and broadens the metallicity distribution function at all radii. 
}

\maketitle

\section{Introduction}

The metal content of stars holds valuable information about the state of the Galaxy at the time and location of their birth. The goal of Galactic archeology is to combine this information with stellar ages and kinematics in order to determine the formation history of the Galaxy (e.g., Chiappini et al. 2001, 2003). In particular, decades of study have sought to understand the chemo-dynamical properties of solar neighbourhood stars: the flat age-metallicity relation, and the fact that both old and young stars can be found at both high and low metallicities. Independent ring models that assume stars born at a particular radius remain there indefinitely predict that younger stars should always form from more metal enriched gas. Therefore a tightly correlated age-metallicity relation with a low degree of scatter is expected, which is not consistent with observations.

However, in recent years it has been shown that stars may change angular momentum at a co-rotation resonance of non-axisymmetric structure, such as spiral structure, which results in the radial shuffling of stars throughout the disc. This ``radial migration'' can complicate the relation between age and metallicity at a given radius, and has been speculated to affect the evolution of the radial metallicity gradient (e.g., Schoenrich \& Binney 2009, Minchev et al. 2012, 2014). However, how and to what extent radial migration drives the global radial metal distribution in the disc is still unknown, owing partially to our lack of knowledge of the past radial metallicity gradient. One of the aims of this proceedings is to summarise the results of our studies on the chemo-dynamical evolution of a spiral disc simulated with numerical methods. 

\begin{figure*}
\includegraphics[scale=0.17]{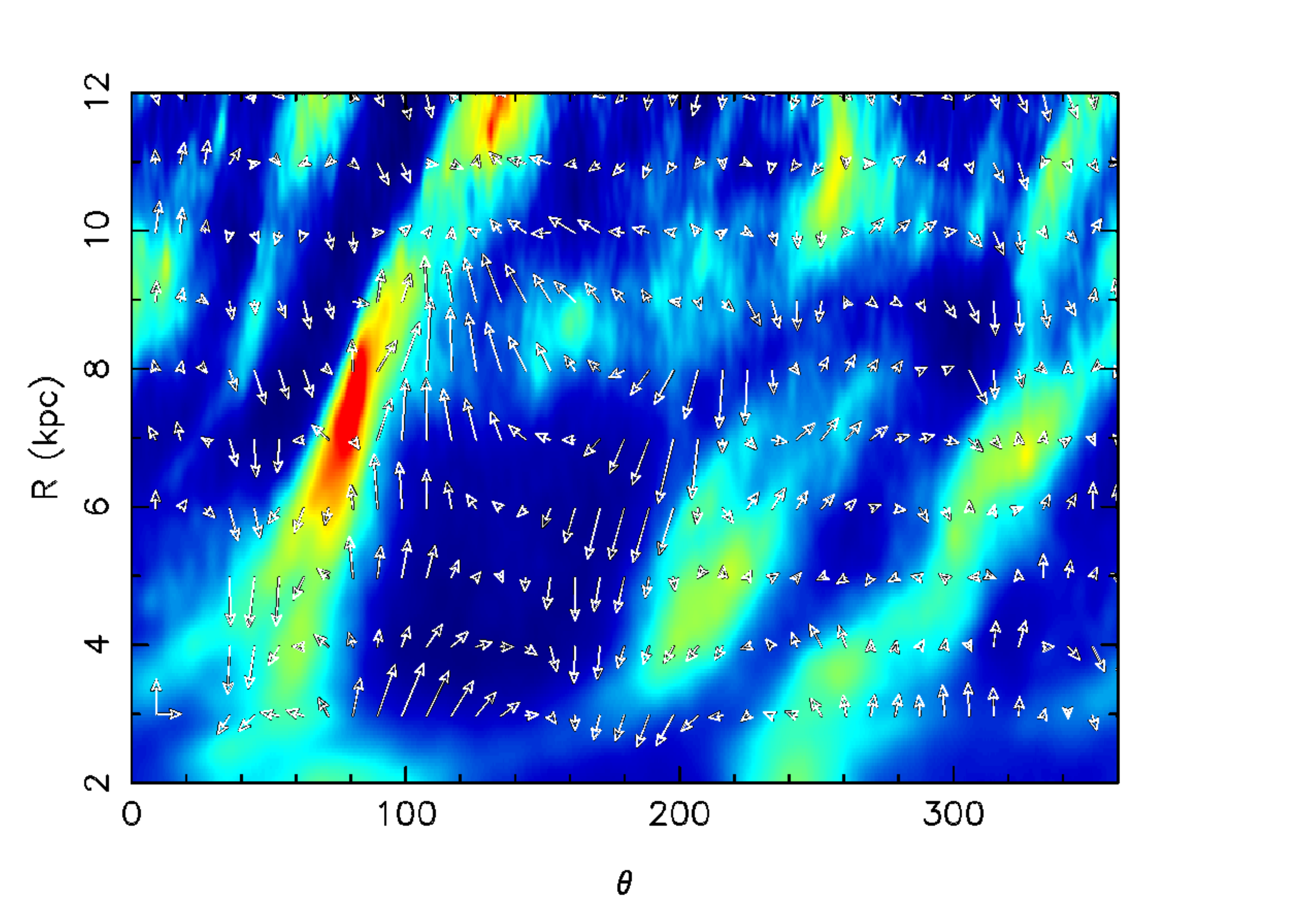}
\hspace{-12.0mm}
\includegraphics[scale=0.61]{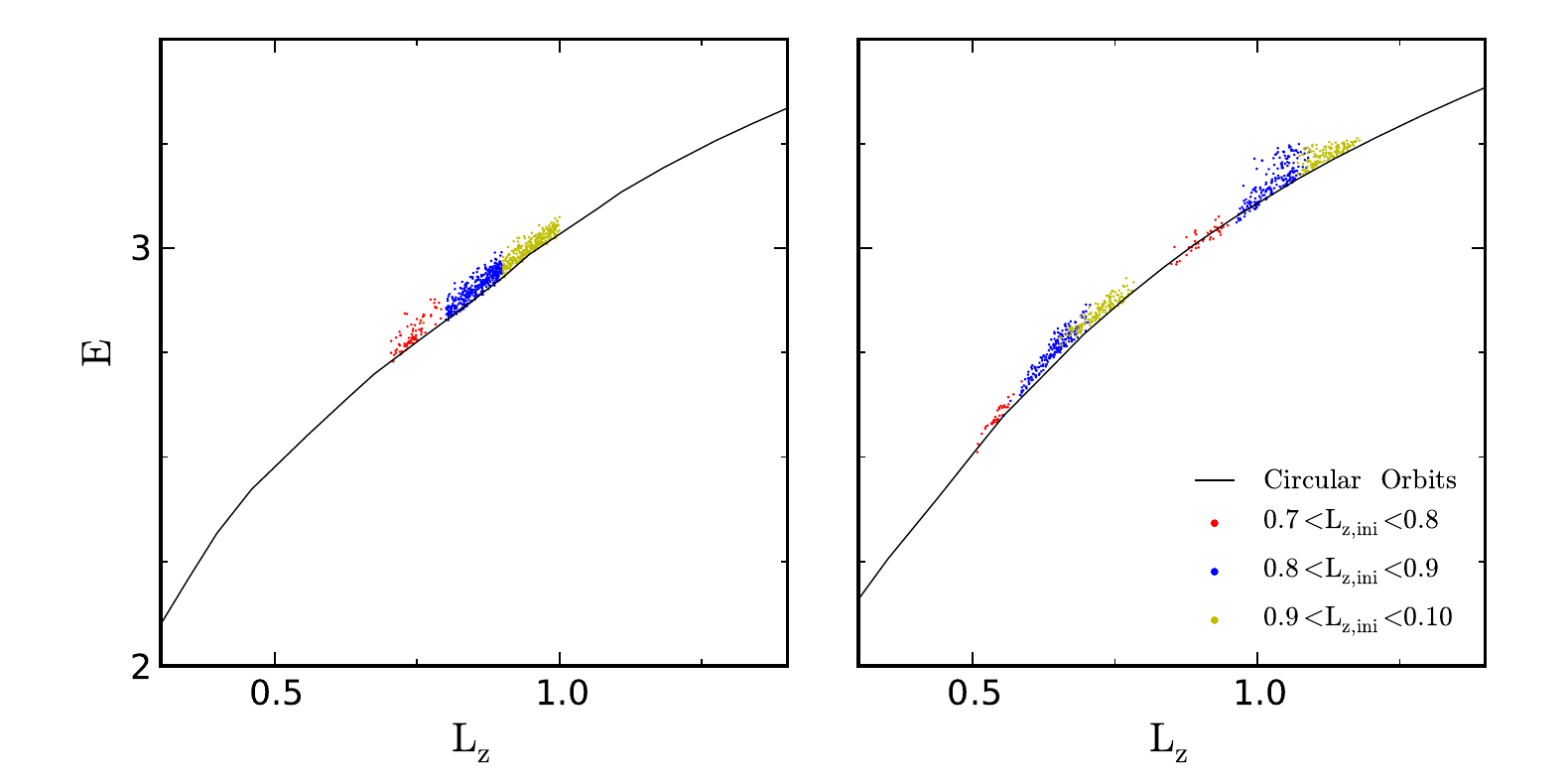}
\caption{\emph{Left:} Stellar over-density map in polar coordinates of a $N$-body simulation with transient, co-rotating spiral arms. White arrows indicate the mean peculiar velocity vector (mean rotation subtracted). Rotation is from right to left. \emph{Middle} and \emph{right:} The locus of three groups of star particles in orbital energy-angular momentum space, selected in different angular momentum ranges from the snapshot in the left panel (initial time) and a 'final' time 100 Myr later (see text for details). The circular orbit line is marked by the black curve. Units are arbitrary.}
\label{f1}
\end{figure*}

The mechanism of radial migration from spiral arms depends on the nature of the spiral arms themselves. In classic density wave theory, the spiral arms are assumed to be rigidly rotating wave crests in the stellar component. In such a theory, there is one special radius at which the spiral arms and stars rotate at the same speed - and at which radial migration can occur: the co-rotation radius. Later, it was shown by Sellwood \& Binney (2002) that spiral arms need necessarily be transient features in order for the effects of radial migration to be long lasting - else migrated stars are returned to their initial angular momentum value on horseshoe orbits. Recently, it has been shown that spiral arms in simulations may be transient, co-rotating features that, because of their ubiquitous co-rotation resonances, cause more radial migration that density waves. It is therefore important to determine the nature of spiral arms are present in the Milky Way.

In this proceedings, {\bf we summarise our analysis of spiral arms in simulations of isolated Milky-Way sized galaxies, and show how radial migration affects the metal distribution. We demonstrate that the spiral-driven peculiar motions provide an observational test of different spiral arm models.}

\section{Simulations}

We present results from a simulation of a Milky-Way sized disc performed with the $N$-body/Smoothed Particle Hydrodynamics (SPH) code GCD+ (Kawata et al. 2013). The simulation consists of a static spherical dark matter halo of mass $M_{DM} = 2.5 \times 10^{12}$ $\rm M_{\odot}$ with a NFW profile (Navarro et al. 1997), and a stellar and a gas disc. The stellar and gas discs have masses $M_* = 4 \times 10^{10}$ $\rm M_{\odot}$ and $M_g = 1 \times 10^{10}$ $\rm M_{\odot}$, and are modelled with $N=4 \times 10^6$ and $N=1 \times 10^6 $ particles respectively. The discs are set up in vertical equilibrium following the method described in Springel et al. (2005). The radial density profiles are set to be exponential with scale lengths $R_* = 2.5$ kpc and $R_g = 4.0$ kpc, respectively. For further details of the simulation, see Kawata et al. (2014) and Hunt et al. (2015).

We run the simulation up to $\sim 1.5$ Gyr, before the computational limitations of the star formation scheme kick in. We focus on the simulation period of $0.5$ - $1.0$ Gyr, by which time a bar of length $\sim 4$ kpc has already formed, as well as a 2-3 armed spiral structure.

\section{Results}

\subsection{Radial migration in transient, co-rotating spiral arms}

The left panel of Fig. \ref{f1} shows a snapshot of the stellar surface over-density at $t=0.5$ Gyr in cylindrical polar coordinates. The main spiral feature is clearly identified within a radial range of 5-10 kpc and an azimuthal range of 50-100$^\circ$. Over-plotted in white arrows are the mean peculiar velocity (mean rotation subtracted) velocity vectors. The systematic streaming motion on both trailing (right side) and leading (left side) side of the spiral arm is strikingly clear: on the trailing side, the tangential force of the spiral arm applies a torque to star particles which gain angular momentum and move outward, and likewise stars on the leading side are torqued radially inward and lose angular momentum. The tangential force is continuously applied to the stars until the spiral arm disrupts (about 100 Myr, see Grand et al. 2012ab). This is because the spiral {\bf stellar density enhancement} rotates at the same speed as the stars at every radius, and therefore the migrating star is always in the vicinity of the spiral as it winds up with the differential rotation. The continual angular momentum gains/loses of these migrator stars manifests as streaming motions along the both sides of the spiral arm. This is qualitatively different from the radial migration that occurs around a density wave-like spiral arm, which is localised to the single co-rotation radius and much less pronounced.

\begin{figure}
\centering
\includegraphics[scale=0.23, trim={1cm, 4.3cm, 0cm, 3cm},clip]{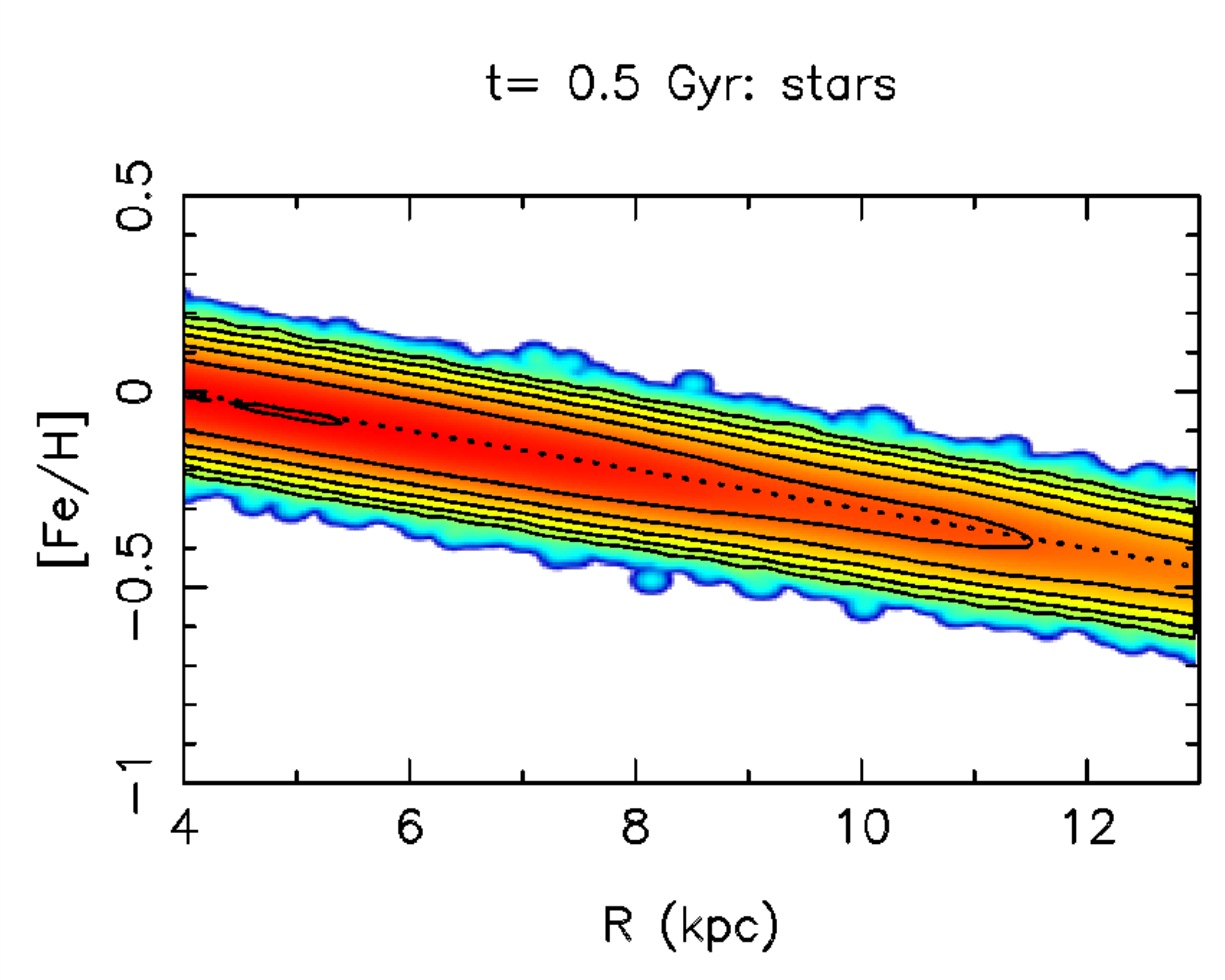} \\
\includegraphics[scale=0.23, trim={1cm, 0.5cm, 0cm, 3cm},clip]{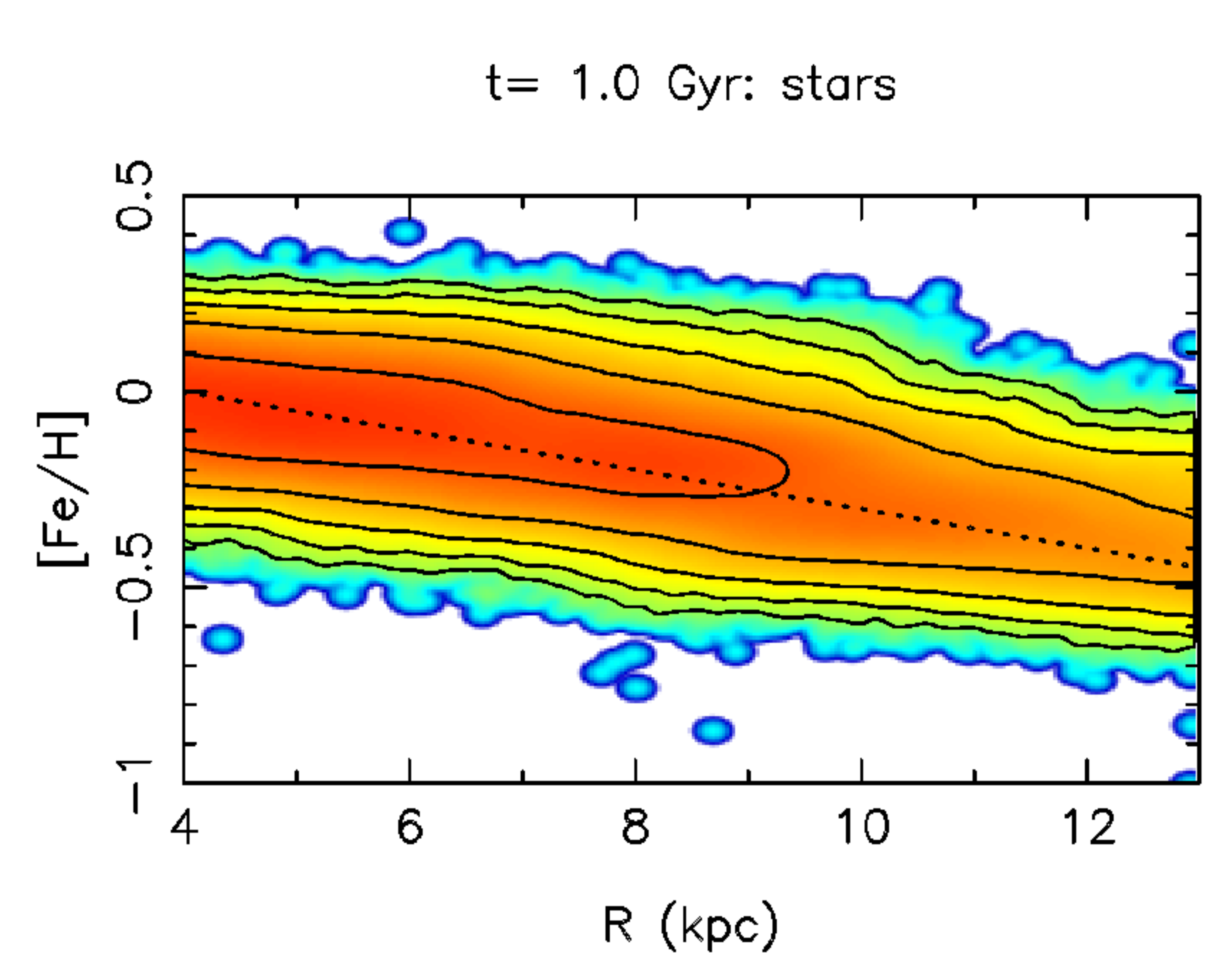} \\
\caption{The radial metallicity distribution of star particles between $0.5$ (top) and $1.0$ (bottom) Gyr. The initial radial metallicity gradient is marked by the dotted black line.}
\label{f2}
\end{figure}

The middle and right panels of Fig. \ref{f1} show the locations of three groups of star particles in energy-angular momentum space at two different times, separated by 100 Myr. The selection criteria of a particle group are as follows: a) select all stars in a given angular momentum range (indicated in the right panel), b) rank these particles in order of their angular momentum change between the two times, c) select the top and bottom percentile of star particles ($\sim$ 500 particles). The middle panel of Fig. \ref{f1} shows that the selected star particles are initially close to the circular orbit line in well defined angular momentum ranges. After 100 Myr of evolution, the selected particles have radially migrated, both increasing and decreasing their angular momentum and orbital energy such that at the second time they are still very close to the circular orbit line (right panel of Fig. \ref{f1}). This behaviour is in agreement with what is expected to occur at co-rotation resonances, and is further indication that the {\bf density enhancement associated with the} spiral arms in $N$-body simulations are transient and co-rotating. It also indicates that spiral arms can cause a lot of radial migration without significantly heating stellar orbits kinematically (as shown in Grand et al. 2012ab).

\subsection{Impact of radial migration on the Radial Metallicity Distribution}

To investigate the impact of radial migration on the radial metallicity distribution (RMD), we examine the time period between 0.5 and 1.0 Gyr. For our purpose, the relevant task is to compute how the RMD changes from 0.5 to 1.0 Gyr, therefore we ``tag'' each particle with a metal value at $t=0.5$ Gyr such that we construct a RMD with a negative radial metallicity gradient equal to $-0.05$ dex kpc$^{-1}$, and a Gaussian metallicity distribution function (MDF) of dispersion $0.05$ dex centred on the mean metallicity at each radius (top panel of Fig. \ref{f2}). The effect of radial migration is then illustrated from the examination of the bottom panel of Fig. \ref{f2}, which shows the RMD at $t=1.0$ Gyr of the same star particles that were present at $t=0.5$ Gyr (i.e., no stars born between these times are considered). {\bf Owing to the one-to-one exchange of particles at each radius, radial migration acts to maintain the negative radial metallicity gradient and broaden the MDF at every radius (Grand et al. 2015a). This result is in contrast with Kubryk et al. (2013), who argue that radial migration can flatten the radial metallicity gradient.} However, {\bf the strong bar formed in Kubryk et al. (2013) is larger than what is observed in the Milky Way, and may cause bulk mass and angular momentum transport (especially during stages of bar growth) that can flatten the gradient.} 

{\bf We note that the simulation presented here is run for only 0.5 Gyr, however we have recently run more $N$-body simulations of barred galaxies up to 8 Gyr (about the age of the Milky Way thin disc), and confirmed this result.}

\subsection{Distinguishing transient, co-rotating spiral arms from density waves}

As mentioned above, the peculiar velocity fields around transient, co-rotating spiral arms are expected to be different from those of density waves. Therefore, scrutiny of the peculiar velocity fields in a suite of simulations with different types of spiral structure provides observational predictions, which can be tested with current Galactic surveys such as APOGEE-RC and RAVE (and in the future, data from the \emph{Gaia} mission). To this end, we analyse a set of simulations comprising the simulation presented above, a further two $N$-body simulations which exhibit spiral structure only (with $\sim2$ and $\sim 6$ arms) and two test particle simulations with {density wave-like} spiral structure only. 

In each simulation, we choose an observer position located behind a spiral arm at approximately 4 kpc distance from the peak spiral over-density in the direction $l=90$, where $l$ is the Galactic longitude. We focus on a square patch in the plane of disc with side length $8$ kpc, and divide it into a grid of square bins of length 800 pc each. In each bin, we calculate the \emph{peculiar} velocity components in the radial and azimuthal directions (defined as any deviation from the mean velocity) of all stars within 250 pc of the midplane, and project along the LOS to obtain the peculiar LOS velocity field. To quantify the fluctuations in LOS peculiar velocity, we take a two-dimensional Fourier transform of this grid of velocity values, and take a one-dimensional azimuthal average of the power that falls in radial annuli in $k$-space, i.e., $k = \sqrt{k_x^2+k_y^2}$. 

\begin{figure}
\centering
\includegraphics[scale=0.45]{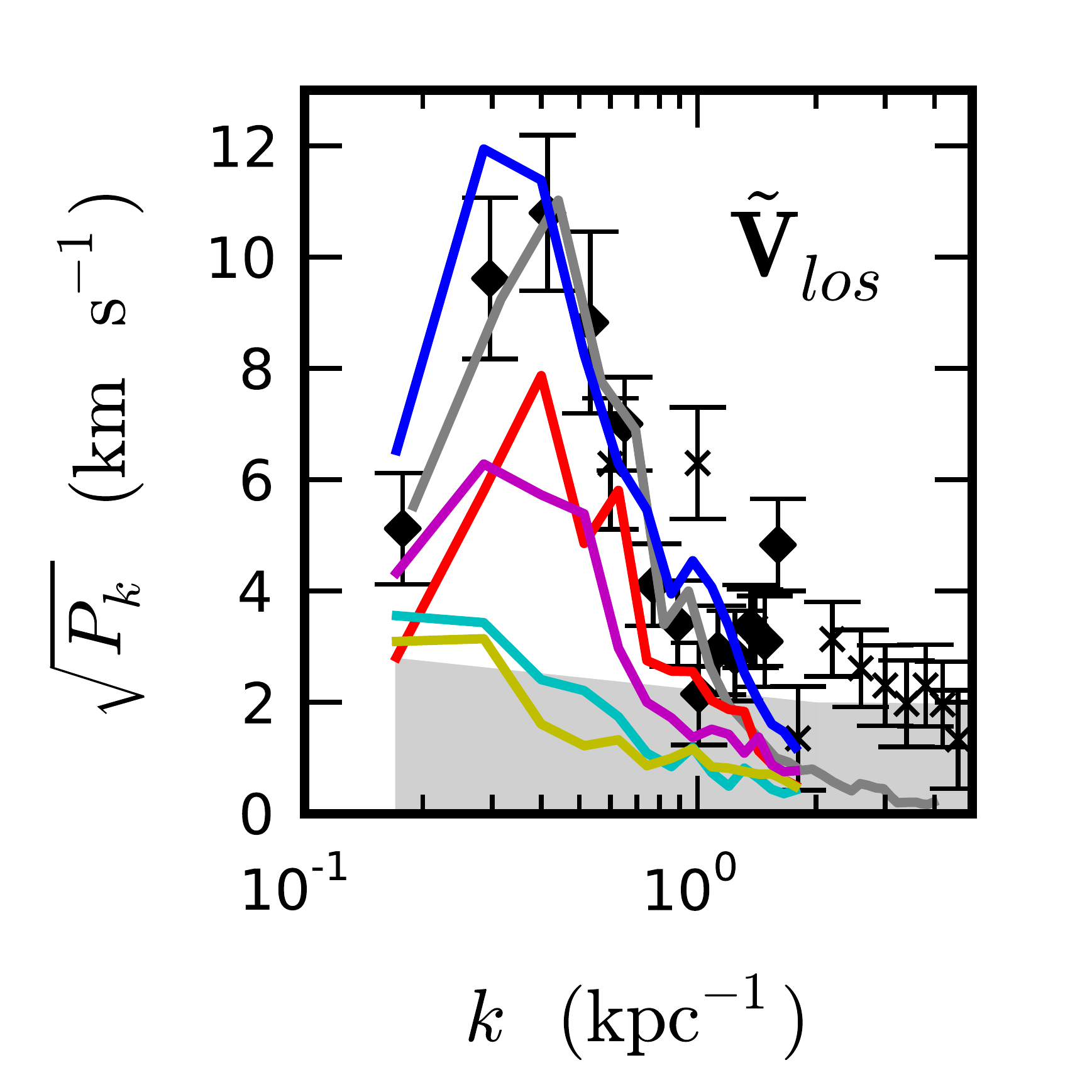}
\caption{The 1D azimuthally averaged power spectrum of the peculiar LOS velocity field for different simulations. Data points from APOGEE-RC and RAVE are given by the black diamonds and crosses, respectively. The grey shaded region indicates the $2$-$\sigma$ noise level of the data.}
\label{f3}
\end{figure}

The results are shown in Fig. \ref{f3}. The blue curve shows the power spectrum of our fiducial model, which peaks at length scales of $k\sim 0.3$ - $0.4$ $\rm kpc ^{-1}$. The shape and location of the peak is consistent with the APOGEE-RC data (black diamonds) and RAVE data (black crosses). The grey curve represents the power spectrum from the bar-only model of Bovy et al. (2015), which fits very well and indicates that the peculiar LOS velocity fields may be driven by the bar. However, the power spectrum of the transient, co-rotating spiral-only $N$-body models (red and magenta curves) show that the location of the observed peak feature can be reproduced by transient, co-rotating spiral arms. We note that the peak amplitude is sensitive to the spiral arm strength, therefore it would be possible to find a transient, co-rotating spiral arm model in which the spiral arms are strong enough to induce the amplitude of the power spectrum. In contrast, the two density wave-like spiral models (green and yellow) exhibit a power spectrum that rises continually as length scales become larger. This is inconsistent with the data, and indicates that density wave-like spiral arms cannot explain the observed peculiar motion in the Milky Way (for more details, see Grand et al. 2015b).

\section{Conclusion}

The main results of our studies are as follows:

\begin{itemize}
\item{} We find that the spiral arms that form in our simulations are transient, recurrent features that co-rotate (Grand et al. 2012ab) and wind up (Grand et al. 2013). 
\item{} The co-rotating nature of the spiral features {\bf allows stars to remain in the vicinity of the over-density while continually gaining (losing) angular momentum on the trailing (leading) side of the spiral} (Grand et al. 2014).
\item{} Migrated stars do not increase their random orbital energy, and therefore remain kinematically cool.
\item{} Radial migration around transient, co-rotating spiral arms does not change the radial metallicity gradient, and broadens the MDF at every radius.
\item{} Peculiar motions associated with radial migration around transient, co-rotating spiral arms appear to be qualitatively consistent with observational data for the Milky Way, whereas those associated with density wave-like spirals are not (Grand et al. 2015b, Monari et al. in prep).
\end{itemize}


\acknowledgements
The authors thank the organisers of the WEH seminar, ``Reconstructing the Milky Way's History''. R.G. acknowledges support by the DFG Research Centre SFB-881 ``The Milky Way System'' through project A1.


%
%

\end{document}